\pdfoutput=1
\documentclass[twocolumn,superscriptaddress,aps,preprintnumbers,amsmath,amssymb,pral]{revtex4-1}
\usepackage{graphicx}
\usepackage{bm}
\usepackage{color}
\usepackage{amsmath,amssymb}	
\usepackage{hyperref}
\usepackage{setspace}
\usepackage{longtable}
\usepackage{booktabs}
\usepackage{comment}
\usepackage{array}
\usepackage{cancel}
\usepackage{enumitem}
\def\slashchar#1{\setbox0=\hbox{$#1$} 
\dimen0=\wd0 
\setbox1=\hbox{/} \dimen1=\wd1 
\ifdim\dimen0>\dimen1 
\rlap{\hbox to \dimen0{\hfil/\hfil}} 
#1 
\else 
\rlap{\hbox to \dimen1{\hfil$#1$\hfil}} 
/ 
\fi}

\def\beq{\begin{eqnarray}}
\def\eeq{\end{eqnarray}}

\begin{document}
\newcolumntype{Y}{>{\centering\arraybackslash}p{23pt}} 


\preprint{IPMU22-0009}

\title{QCD Axion Search with ILC Beam Facility 
}

\author{Hajime Fukuda}
\affiliation{Theoretical Physics Group, Lawrence Berkeley National Laboratory,
 CA 94720, USA}
\affiliation{Berkeley Center for Theoretical Physics, Department of Physics,\\
 University of California, Berkeley, CA 94720, USA}

\author{Hidetoshi Otono} 
\affiliation{Research Center for Advanced Particle Physics, Kyushu University, Fukuoka 819-0395, Japan}

\author{Satoshi Shirai}
\affiliation{Kavli IPMU (WPI), UTIAS, The University of Tokyo, Kashiwa, Chiba 277-8583, Japan}

\date{\today}
\begin{abstract}
One of the most promising methods to search for axions is a light-shining-through-walls (LSW) experiment.
In this work, we discuss the possibility of performing an LSW experiment at the ILC experiment, where photon beams are generated for positron production.
The photon beam is energetic and intense; the energy is of order MeV and the number of photons is about $10^{24}$ per year.
Due to the high energy and intensity, this LSW experiment can reveal the parameter region of the axion unexplored by previous ground-based experiments.
\end{abstract}

\maketitle

\section{Introduction}
\label{sec:introduction}
The International Linear Collider (ILC)\,\cite{Behnke:2013xla} is a future electron-positron linear collider. 
Unlike a circular collider, storage and recycling of beam particles are not possible in a linear collider.
To obtain a large luminosity, we need a positron source of strong intensity at the ILC.
It is, however, challenging to develop a high-intensity positron source. 
One promising proposal is to use the main electron beam. 
In this proposal, the electron beams pass into a helical undulator before the interacting point (IP).
At the undulator, spatially oscillating magnetic fields are imposed and an intense photon beam is emitted from the electron beam. 
The photon beam then goes through a thin target, generating $e^+e^-$ pairs\,\cite{Riemann:2020ytg}.
In order to generate abundant positrons at the ILC, the photon beam from the undulator must be intense and energetic enough.
Indeed, it is the strongest MeV photon beam available on the ground up to the present.
However, after the electron-positron pair creation, the photon is not used anymore but dumped in the current design.

In this work, we propose to use this photon beam for a search for new light particles, in particular, axions. 
An axion is a pseudo Nambu-Goldstone boson associated with the spontaneous breaking of a global $\mathrm{U}(1)_{\mathrm{PQ}}$ symmetry, which is anomalous on the QCD sector. It is introduced to solve the strong CP problem in the standard model (SM)\,\cite{Peccei:1977hh,Peccei:1977ur,Weinberg:1977ma,Wilczek:1977pj}, where the $\theta$ angle in the QCD is unnaturally small\,\cite{Tanabashi:2018oca}. 
It is one of the most important targets for physics beyond the SM. 
Due to various experimental constraints, the scale of $\mathrm{U}(1)_{\mathrm{PQ}}$ symmetry breaking should be much greater than the electroweak scale.
Prime examples of such invisible axion models are KSVZ \cite{Kim:1979if,Shifman:1979if} and DFSZ \cite{Dine:1981rt,Zhitnitsky:1980tq} models.
With the color anomaly of the $\mathrm{U}(1)_{\mathrm{PQ}}$ symmetry, an axion couples to gluons. 
After the QCD phase transition, the axion-gluon coupling induces mixings between the axion and mesons, resulting in the axion mass. 
The axion also couples to photons from the UV physics and the mixing with mesons.

In the presence of a magnetic field, the axion-photon coupling induces the mixing between an axion and a photon; a photon can oscillate into an axion and vice versa\,\cite{Sikivie:1983ip}. 
A type of experiment to use this oscillation to detect an axion is called a light-shining-through-walls (LSW) experiment\,\cite{Anselm:1985obz,VanBibber:1987rq}. 
In LSW experiments, a photon detector is placed far away from a photon beam source in the beam direction.
Between the detector and the beam source, a thick wall is placed to shield the photon beam. 
Magnetic fields perpendicular to the beam direction are imposed on the cavity. 
If the mixing between the axion and photon is large enough, the photon from the source can convert into an axion before the wall, and the axion penetrates the wall.
The axion after the wall may oscillate back to a photon by the magnetic field again. 
The re-converted photon is detected at the detector. 
In this way, LSW experiments can detect axion particles.
The LSW experiment is one of the most promising ways to search for axions and a number of experiments have been performed\,\cite{ALPS:2009des,Bahre:2013ywa,Ruoso:1992nx,Cameron:1993mr,Robilliard:2007bq,GammeVT-969:2007pci,Afanasev:2008jt,OSQAR:2015qdv,Inada:2016jzh}. 

Our proposal is to perform an LSW experiment at the ILC by using the MeV photon beam for positron production.
The photon dump in the ILC plays the role of the thick wall in an LSW experiment. 
The dump can be placed about $2\,\mathrm{km}$ away from the undulator\,\cite{positronSource}. 
We propose to install magnets in the vacancy before and after the dump to induce the photon-axion mixing.
The converted ``axion beam" is invisible and goes through the dump.
The axions oscillate back to a photon, being detected at the photon detector.
As the photon beam is well collimated and the opening angle is of order of the inverse of the Lorentz factor of the main electron beam, $O(10^{-6})\,\mathrm{rad}$, the beam spread is less than cm even at distances of more than 1 km.
This makes it possible to efficiently impose a magnetic field to the axion/photon beam over several kilometers or even more.
Depending on the details of the experiment, such as the tunnel structure, 
it is possible to apply a magnetic field inside the ILC tunnel and its extension, 
or it is also possible to apply a magnetic field on the ground to the beam coming through the ground.
We can thus construct an LSW experiment using the ILC facilities.

This letter is organized as follows; in Sec.\,\ref{sec:physics}, we discuss the conversion probability between an axion and a photon and show the advantage to use the MeV photon at the ILC for the LSW experiment. 
We discuss experimental setups in Sec.\,\ref{sec:experiment} and show the expected sensitivities in Sec.\,\ref{sec:result}. 
Finally, Sec.\,\ref{sec:conclusion} is devoted to the conclusion and discussion.

\section{Axion-photon Conversion}
\label{sec:physics}
Here, let us discuss the axion-photon conversion probability.
In the interaction picture, the axion-photon conversion probability in the presence of the magnetic field is
\begin{align}
    P(a \to \gamma) = \frac{\left|\langle a(p)|T \exp(- i\int H_I dt)|\gamma(k)\rangle\right|^2}{\langle a(p)|a(p)\rangle\langle \gamma(k)|\gamma(k)\rangle},
\end{align}
where $p(k)$ is the four momentum of the axion (photon) and $H_I$ is the interaction Hamiltonian,
\begin{align}
    H_I = \int d^3 x \frac14 g_{a\gamma\gamma} a F_{\mu\nu} \tilde{F}^{\mu\nu}.
\end{align}
With the normalization of the annihilation and creation operators
\begin{align}
    [a_i(p), a_j^\dagger(p')] = (2\pi)^3 2p^0 \delta_{ij}\delta^3(p - p'),
\end{align}
where $i$ and $j$ is the species of a particle, 
the denominator is
\begin{align}
    \langle a(p)|a(p)\rangle\langle \gamma(k)|\gamma(k)\rangle = 2p^0  2k^0 V^2,
\end{align}
where  $V = (2\pi)^3 \delta^3(0)$ is the volume of the system.

Let us assume the axion and photon are moving in $z$ direction and the magnetic field $\vec{B} = \vec{B}(z)  $ depends only on $z$. 
In the lowest order approximation, the numerator is
\begin{align}
    \langle a(p)|&T \exp(- i\int H_I dt)|\gamma(k)\rangle \simeq \nonumber\\
    & -ig_{a\gamma\gamma} \int d^4 x B_i(z) \langle a(p)| a F_{0i} |\gamma(k)\rangle.
\end{align}
For $p^0 = k^0 \equiv \omega$, it reduces to
\begin{align}
    g_{a\gamma\gamma} \omega S T \int dz e^{iqz} \vec{B}(z)\cdot \vec{e}(k),
\end{align}
where $S$ is the area in $xy$ direction, $T$ is the total time for the conversion, $q = \omega - \sqrt{\omega^2 - m_a^2}$ is the momentum transfer and $\vec{e}$ is the polarization vector of the final state photon. For high energy axions, $\omega \gg m_a$, $T$ is equivalent to the length of the system in $z$ direction, $L$.

Combining the numerator and the denominator, the conversion probability is
\begin{align}
    P(a \to \gamma;\omega) &\simeq \frac{\omega^2 g_{a\gamma\gamma}^2 S^2 L^2 \left|\int dz e^{iqz} \vec{B}(z)\cdot \vec{e}(k)\right|^2}{4\omega^2 V^2} \nonumber \\
    & = \frac{g_{a\gamma\gamma}^2}{4} \left|\int dz e^{iqz} \vec{B}(z)\cdot \vec{e}(k)\right|^2.
    \label{eq:conversion}
\end{align}
For a constant magnetic field, $B_x(z) = B_0$, the formula reduces to
\begin{align}
    P(a \to \gamma;\omega) = \frac14 g_{a\gamma\gamma}^2 B_0^2 L^2 \left[\frac{\sin (qL / 2)}{qL / 2}\right]^2.
\end{align}
For $\omega \gg m_a$, the momentum transfer $q$ is
\begin{align}
    q\simeq \frac{m_a^2}{2\omega} = \left(10\,\mathrm{km}\right)^{-1} \left(\frac{m_a}{10^{-2}\,\mathrm{eV}}\right)^2 \left(\frac{2.5\,\mathrm{MeV}}{\omega}\right).
\end{align}
For the smaller value of the momentum transfer $q \ll 1/L$, the larger $L$ can enhance the conversion rate.
However, for $q L \gtrsim 1$, the conversion rate gets smaller and suppressed by $q^{-2}$.
In terms of the axion mass, the conversion rate becomes smaller for $m_a \gg \sqrt{\omega /L}\sim 10^{-4}~\mathrm{eV} (\omega/1~\mathrm{eV})^{1/2}  (L/1~\mathrm{m})^{-1/2} $.
Therefore, LSW experiments using visible laser beams, $\omega = O(1)~\mathrm{eV}$, do not have a good sensitivity to the high mass QCD axion.

Several ways are possible to improve the sensitivity for the heavier mass region, $m_a = {O}(0.1\mbox-1)\,\mathrm{eV}$.
First, one can use photons with higher energy. This suppresses the momentum transfer and we may have longer $L\lesssim 1/q$.
As we will see later, the energy of the undulator photon at the ILC can be $O(10)~\mathrm{MeV}$.
We can maintain the sensitivity of sub-eV QCD axion even for $L \sim \mathrm{km}$.
Second, we may adopt magnetic fields wiggled spatially with a period of $O(1/q)$ to improve the heavier-mass sensitivity more.
In this case, the integral of the conversion rate \eqref{eq:conversion} is no longer canceled for $q L \sim 1$ and the conversion rate can be enhanced for the larger size $L$.
Finally, we may also give an effective mass $\omega_p$ to the photon to suppress the momentum transfer\,\cite{Inoue:2002qy,CAST_He4,CAST_He3}. This can be done by filling the whole conversion volume with, for example, helium gas. However, it is not obvious whether a small plasma frequency of order $\omega_p \sim m_a = {O}(0.1\mbox-1)\,\mathrm{eV}$ affects the coherent oscillation between a photon and an axion as energetic as $O(10)\,\mathrm{MeV}$. Moreover, the gas would absorb photons and a longer conversion length could not be used. Thus, we do not adopt this method in this work.

\section{Experimental Setup}
\label{sec:experiment}
\begin{figure}[t]
{\includegraphics[width=0.5\textwidth]{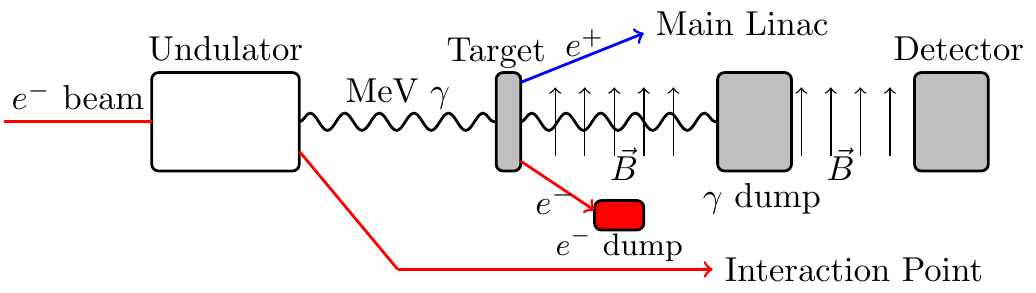}} 
\caption{
A schematic of the positron production system at the ILC based on undulator photon, and proposed LSW experiment for the axion detection.
}
\label{fig:setup}
\end{figure}
The schematic picture of our proposal is shown in Fig.\,\ref{fig:setup}, assuming the undulator-based positron source\,\cite{Adolphsen:2013kya}. The main electron beam is passed through a helical undulator to produce gamma rays with energies around 10 MeV. Subsequently, the photons are collided with a thin target to produce positrons.
We propose to use the photon beam after the target as the source of an LSW experiment for the axion search.
In the present design of the ILC, the photon beam pipe will be installed about $1.5\,\mathrm{m}$ away from the electron main beam pipe. We thus may install magnets and apply a magnetic field on the photon beam in the section between the photon dump and the target.
The magnetic field induces the photon-axion mixing and a fraction of photons are converted into axions, passing through the photon dump. 
We impose a magnetic field behind the dump to turn the axion back into a photon again, which is then captured by the detector.

The advantage of performing the LSW experiment by the ILC photon beam is the high photon energy; as discussed in Sec.\,\ref{sec:physics}, the sensitivity to heavier axion mass regions does not decrease even if a magnetic field is applied over a long distance. 
To improve the sensitivity to heavier axions more, we also propose to make the magnetic field spatially periodic.

\subsection*{Undulator Photon}
Let us review the photon beam system at the ILC\,\cite{Adolphsen:2013kya}.
For the ILC250 main electron beam, the pulse repetition rate is $5~\mathrm{Hz}$ and each pulse contains 
1312 bunches containing $\sim 10^{12}$ electrons with a bunch separation of $554\,\mathrm{ns}$ and a bunch length $0.3\,\mathrm{mm}$.
In the design proposal, 
the undulator period is $\lambda = 11.6\,\mathrm{mm}$ and the undulator parameter is $K \equiv e B \lambda / 2 \pi m_e = O(1)$.
The typical energy of the undulator photon is $\omega \sim  4\pi \gamma_e^2/(1 + K^2) \lambda$ with $\gamma_e$ is the Lorentz factor of the electrons.
The opening angle of the beam is around $1/\gamma_e = O(10^{-6})$.
The photon beam power is $O(10\mbox{-}100)\,\mathrm{kW}$. 

The undulator photons are collided with a thin target to generate positrons.
A few percent of the photons from the undulator are absorbed in the thin target and the rest passes through it.
The remaining photon beams are absorbed in a photon dump system.
The photon dump is proposed to be placed about $2\,\mathrm{km}$ away from the undulator, as the divergence of the photon beam can relax the temperature rises of the beam window\,\cite{positronSource}.

\subsection*{Photon Detector}
The location of the re-conversion cavity and the photon detector depends on the details of the ILC tunnel and beam designs.
The ILC tunnel will not be straight as it follows the horizontal plane of the earth and there is a non-zero crossing angle at the collision point.
There are several possible locations for the re-conversion cavity.
Here we discuss the possibility of installing the cavity and detector in a different tunnel than the ILC main tunnel,
which turns out to be an underground low-level counting experiment. 

The signal of the experiment is a single photon with an energy of around 10~MeV, while cosmic rays could be the main sources of background: either neutrons from the negative muon capture or direct ionization in the detector. 
Radionuclides in the detector environment could not be the background, since photon energy from the radionuclides would not be more than 3~MeV.
Ref.\,\cite{b8e9aa681f8b4f3291a77ffb2650033d} shows that a simple 76~mm-diameter and 76~mm-long High Purity Germanium (HPGe) Radiation Detectors located in an underground laboratory at 20~m~w.e. recorded 1 count/day/MeV around 10~MeV if additional detectors for the anticosmic suppression are installed (see Fig.~2.15 in the reference). 
Since the location of the photon detector for the experiment is supposed to be at about 200~m below the surface of the ground, 
the muon flux could be attenuated by a factor of $10^{-3}$ (see Figs.~1.3 and 2.9 in the reference).
As discussed before, the photon beams have a pulsed structure.
The timing information can be used to suppress background.
Time resolution of HPGe is expected to be 3-4 nanoseconds \cite{509}, which provides a reduction factor of $10^{-4}$.
Consequently, the background is estimated to be negligible even for 10 years running of the ILC250. 

Moreover, we can also utilize the angular information of the photon.
The photon from the axion should be collimated in $O(1)\,\mathrm{cm}$ radius at the detector location.
Various position and angular sensitive detectors for MeV gamma rays are developed for astrophysical and medical applications.
For example, the AMEGO detector \cite{AMEGO:2019gny} and eASTROGAM \cite{e-ASTROGAM:2016bph,e-ASTROGAM:2017pxr} would provide a position resolution around 1~mm and an angular resolution of around 1~degree at 10~MeV.  
This information can further suppress the background.
In the following discussion, we assume the background can be zero and the detection efficiency of the photon is 100\%.

\section{Expected Sensitivity}
\label{sec:result}

Let us estimate the number of observed photons.
For a photon with an energy $\omega$ and polarized in the direction of a magnetic field, the conversion rate is given by, as we have derived in Sec.\,\ref{sec:physics},
\begin{align}
    P(\gamma \to a;\omega) \simeq \frac{g_{a\gamma\gamma}^2}{4} \left|\int dz e^{iqz} B_1(z)\right|^2,
\end{align}
where $B_1(z)$ is a configuration of the magnetic filed in the photon conversion region. 
Similarly, the re-conversion from an axion to a photon is given by
\begin{align}
    P(a \to \gamma;\omega) \simeq \frac{g_{a\gamma\gamma}^2}{4} \left|\int dz e^{-iqz} B_2(z)\right|^2,
\end{align}
with $B_2(x)$ being a magnetic field in the re-conversion region.

The undulator photons used in this experiment are not monochromatic but have a finite energy spectrum $\mathcal{F}(\omega)$, where the number of photons with energy from $\omega$ to $\omega + d\omega$ is $\mathcal{F}(\omega) d\omega$.  
The final result of the expected number of the photons $\mathcal{N}$ is given by
\begin{align}
    \mathcal{N} = \frac{1}{2} \int d \omega \mathcal{F}(\omega)  P(\gamma \to a;\omega) \times P(a \to \gamma;\omega).
\end{align}
Here, an extra $1/2$ factor appears, as the ILC undulator photon is circularly polarized.
In the following analysis, we adopt the photon spectrum $\mathcal{F}(\omega)$ from Refs.\,\cite{Alharbi:2021okl,Alharbi:2021ctf} and take $\omega > 5\,\mathrm{MeV}$ to suppress the background from the radionuclides.
As for the configuration of the magnetic fields, we consider the uniform and wiggled cases.
For the wiggled magnetic field, we use a square wave with the spatial period $2w_B$, $B(z) = B_0 \times \mathrm{sgn}(\sin(\pi z/w_B))$, for simplicity.

\begin{figure}[t]
{\includegraphics[width=0.5\textwidth]{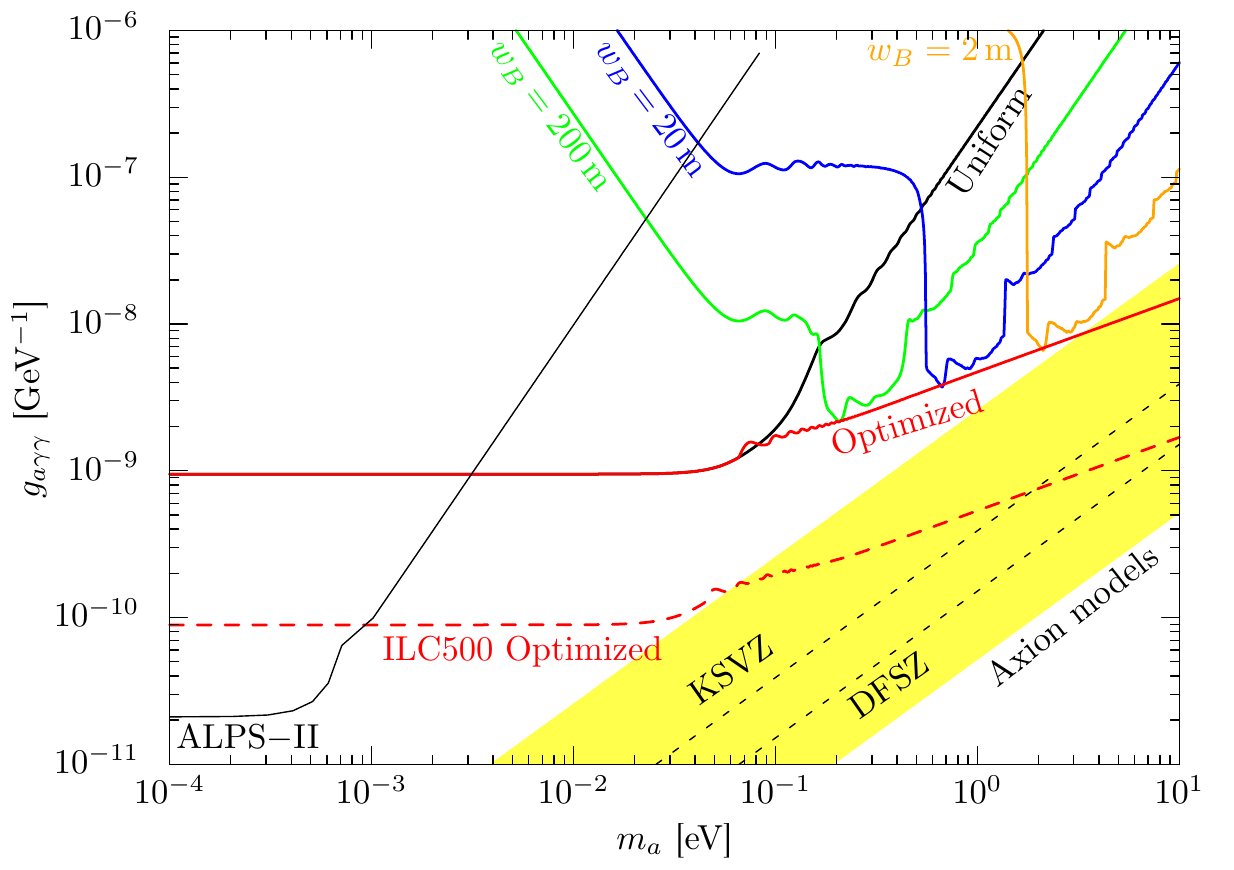}} 
\caption{
The expected sensitivity of the ILC LSW experiment.
The yellow region shows the range of the QCD axion models \cite{DiLuzio:2016sbl,*DiLuzio:2017pfr}.
}
\label{fig:result}
\end{figure}

In Fig.\,\ref{fig:result}, we show the expected detection sensitivity of the QCD axion at 95\% C.L. assuming the background is zero, i.e., $ \mathcal{N}  = 3.0$.
In the solid lines, we show the sensitivity of 10 years running of the ILC250, which produces around $10^{25}$ undulator photons. 
Here we assume that the conversion and re-conversion length are $2\,\mathrm{km}$ and the amplitude of the magnetic field $B_0$ is $1\,\mathrm{T}$,
In addition to the uniform magnetic field, we show the magnetic field wiggled with $w_B = 200, 20$ and $2\,\mathrm{m}$.
In these cases, the sensitivity for the axion mass satisfying $q w_B = O(1)$ can be improved.
Moreover, we also show the case that the value of $w_B$ is optimized for each mass: $w_B  = \pi/q \simeq 2 \pi \bar{\omega}/m_a^2$, where $\bar{\omega}$ is a typical energy of the undulator photon and 7(16)\, MeV for the ILC250(ILC500).
In the dashed line, we represent the optimized  case for the ILC500.
Here we adopt $10\,\mathrm{km}$ conversion and re-conversion distances and  $2\,\mathrm{T}$ magnetic fields.
In this figure, we also show the sensitivity of the ALPS-II\,\cite{Bahre:2013ywa}. 
Compared with the LSW experiments based on optical or infrared lasers, the present ISW experiment at the ILC can probe higher mass axion.

\section{Conclusion and Discussion}
\label{sec:conclusion}
In this work, we have proposed an LSW experiment using an MeV undulator photon beam at the ILC beam facility. 
Our proposal requires to install magnets along the photon beam line and build a cavity with magnets and a detector.
Thanks to the high-energy photon, we will be able to probe high-mass axion parameters which have not been explored by ground-based experiments.

One of the advantages of the present method is that it is highly expandable.
Basically, the greater the length of the magnetic field applied, the better the sensitivity to the axion-photon coupling.
If we adopt a magnetic field wiggled with a period $O(1)~\mathrm{cm} - O(100)~\mathrm{m}$, we can test high-mass axion, which is a well-motivated parameter region as a solution to the strong CP problem.

Of course, how long the magnetic field can be imposed and where the photon detector can be placed strongly depends on the detailed design of the ILC, such as tunnel and beam designs.
Depending on these designs, they may be placed in the ILC tunnel itself or in its extension.
Another possibility is to place the re-conversion cavity and detector on the ground.
After the photon dump, the invisible ``axion beam" goes straight into the rock around the ILC tunnel. 
It will eventually appear on the ground. 
As the photon/axion beam is well collimated, the size of the axion beam spread after can still be small on the ground.
We put magnets in the direction of the axion beam and a detector at the end of the magnets.
It also works as an LSW experiment.

Let us discuss the capability of the ILC LSW experiment to detect other kinds of new particles. 
Another important target at LSW experiments is a dark photon $\gamma'$, a massive vector boson with non-vanishing kinetic mixing $\chi$ with an SM photon. 
With this mixing $\chi$, the interaction and energy eigenstates of the vector bosons are not identical. 
As a result, a photon, the SM interaction eigenstate, oscillates into a dark photon, the hidden interaction eigenstate with a momentum $p$, and vice versa, even without any external fields like magnetic fields.
Thus, LSW experiments constrain the mass and mixing parameters of the dark photon.
The conversion rate is $P(\gamma \leftrightarrow \gamma') \sim \chi^2$ for the dark photon of a mass much less than the photon energy.
In this case, we estimate the LSW experiment with the ILC250 setup may reach $\chi \simeq 5 \times 10^{-7}$.

Similar to the axion search we have discussed, the use of a high-energy photon beam is expected to enhance the sensitivity for heavier dark photon mass regions \,\cite{Inada:2013tx}.
However, the precise upper bound on the dark photon mass is not clear for the following two reasons. 
First, the mass threshold, in other words, $\sqrt{s}$, of the photon creation process is not clear. 
It depends on the detailed design of the undulator. 
Second, for heavier dark photon masses, the different mass eigenstates decohere; a vector boson at the source is not the SM interaction eigenstate but a mixed state of the mass eigenstates. 
To estimate the decoherence effect, we need to take into account how localized wave packets are\,\cite{Akhmedov:2010ua}. We leave the analysis of the dark photon for future work.

\begin{acknowledgments}
We would like to thank Yu Morikawa for helpful discussions on the undulator positron source at the ILC.  
This work is supported in part by JSPS Grant-in-Aid for Scientific Research  17H02878 (S.S.),  18K13535(S.S.), 19H04609(S.S.), 20H01895(S.S.), 20H05860(S.S.) and 21H00067(S.S.) and by World Premier International Research Center Initiative (WPI Initiative), MEXT, Japan (S.S.) by the Director, Office of Science, Office of
High Energy Physics of the U.S. Department of Energy under the
Contract No. DE-AC02-05CH11231 (H.F.).
\end{acknowledgments}

\bibliography{papers}

\begin{thebibliography}{39}%
\makeatletter
\providecommand \@ifxundefined [1]{%
 \@ifx{#1\undefined}
}%
\providecommand \@ifnum [1]{%
 \ifnum #1\expandafter \@firstoftwo
 \else \expandafter \@secondoftwo
 \fi
}%
\providecommand \@ifx [1]{%
 \ifx #1\expandafter \@firstoftwo
 \else \expandafter \@secondoftwo
 \fi
}%
\providecommand \natexlab [1]{#1}%
\providecommand \enquote  [1]{``#1''}%
\providecommand \bibnamefont  [1]{#1}%
\providecommand \bibfnamefont [1]{#1}%
\providecommand \citenamefont [1]{#1}%
\providecommand \href@noop [0]{\@secondoftwo}%
\providecommand \href [0]{\begingroup \@sanitize@url \@href}%
\providecommand \@href[1]{\@@startlink{#1}\@@href}%
\providecommand \@@href[1]{\endgroup#1\@@endlink}%
\providecommand \@sanitize@url [0]{\catcode `\\12\catcode `\$12\catcode
  `\&12\catcode `\#12\catcode `\^12\catcode `\_12\catcode `\%12\relax}%
\providecommand \@@startlink[1]{}%
\providecommand \@@endlink[0]{}%
\providecommand \url  [0]{\begingroup\@sanitize@url \@url }%
\providecommand \@url [1]{\endgroup\@href {#1}{\urlprefix }}%
\providecommand \urlprefix  [0]{URL }%
\providecommand \Eprint [0]{\href }%
\providecommand \doibase [0]{http://dx.doi.org/}%
\providecommand \selectlanguage [0]{\@gobble}%
\providecommand \bibinfo  [0]{\@secondoftwo}%
\providecommand \bibfield  [0]{\@secondoftwo}%
\providecommand \translation [1]{[#1]}%
\providecommand \BibitemOpen [0]{}%
\providecommand \bibitemStop [0]{}%
\providecommand \bibitemNoStop [0]{.\EOS\space}%
\providecommand \EOS [0]{\spacefactor3000\relax}%
\providecommand \BibitemShut  [1]{\csname bibitem#1\endcsname}%
\let\auto@bib@innerbib\@empty
\bibitem [{\citenamefont {Behnke}\ \emph {et~al.}(2013)\citenamefont {Behnke},
  \citenamefont {Brau}, \citenamefont {Foster}, \citenamefont {Fuster},
  \citenamefont {Harrison}, \citenamefont {Paterson}, \citenamefont {Peskin},
  \citenamefont {Stanitzki}, \citenamefont {Walker},\ and\ \citenamefont
  {Yamamoto}}]{Behnke:2013xla}%
  \BibitemOpen
  \bibfield  {author} {\bibinfo {author} {\bibfnamefont {T.}~\bibnamefont
  {Behnke}}, \bibinfo {author} {\bibfnamefont {J.~E.}\ \bibnamefont {Brau}},
  \bibinfo {author} {\bibfnamefont {B.}~\bibnamefont {Foster}}, \bibinfo
  {author} {\bibfnamefont {J.}~\bibnamefont {Fuster}}, \bibinfo {author}
  {\bibfnamefont {M.}~\bibnamefont {Harrison}}, \bibinfo {author}
  {\bibfnamefont {J.~M.}\ \bibnamefont {Paterson}}, \bibinfo {author}
  {\bibfnamefont {M.}~\bibnamefont {Peskin}}, \bibinfo {author} {\bibfnamefont
  {M.}~\bibnamefont {Stanitzki}}, \bibinfo {author} {\bibfnamefont
  {N.}~\bibnamefont {Walker}}, \ and\ \bibinfo {author} {\bibfnamefont
  {H.}~\bibnamefont {Yamamoto}},\ }\href@noop {} {\  (\bibinfo {year}
  {2013})},\ \Eprint {http://arxiv.org/abs/1306.6327} {arXiv:1306.6327
  [physics.acc-ph]} \BibitemShut {NoStop}%
\bibitem [{\citenamefont {Riemann}\ \emph {et~al.}(2020)\citenamefont
  {Riemann}, \citenamefont {Sievers}, \citenamefont {Moortgat-Pick},\ and\
  \citenamefont {Ushakov}}]{Riemann:2020ytg}%
  \BibitemOpen
  \bibfield  {author} {\bibinfo {author} {\bibfnamefont {S.}~\bibnamefont
  {Riemann}}, \bibinfo {author} {\bibfnamefont {P.}~\bibnamefont {Sievers}},
  \bibinfo {author} {\bibfnamefont {G.}~\bibnamefont {Moortgat-Pick}}, \ and\
  \bibinfo {author} {\bibfnamefont {A.}~\bibnamefont {Ushakov}},\ }in\
  \href@noop {} {\emph {\bibinfo {booktitle} {{International Workshop on Future
  Linear Colliders}}}}\ (\bibinfo {year} {2020})\ \Eprint
  {http://arxiv.org/abs/2002.10919} {arXiv:2002.10919 [physics.acc-ph]}
  \BibitemShut {NoStop}%
\bibitem [{\citenamefont {Peccei}\ and\ \citenamefont
  {Quinn}(1977{\natexlab{a}})}]{Peccei:1977hh}%
  \BibitemOpen
  \bibfield  {author} {\bibinfo {author} {\bibfnamefont {R.}~\bibnamefont
  {Peccei}}\ and\ \bibinfo {author} {\bibfnamefont {H.~R.}\ \bibnamefont
  {Quinn}},\ }\href {\doibase 10.1103/PhysRevLett.38.1440} {\bibfield
  {journal} {\bibinfo  {journal} {Phys. Rev. Lett.}\ }\textbf {\bibinfo
  {volume} {38}},\ \bibinfo {pages} {1440} (\bibinfo {year}
  {1977}{\natexlab{a}})}\BibitemShut {NoStop}%
\bibitem [{\citenamefont {Peccei}\ and\ \citenamefont
  {Quinn}(1977{\natexlab{b}})}]{Peccei:1977ur}%
  \BibitemOpen
  \bibfield  {author} {\bibinfo {author} {\bibfnamefont {R.}~\bibnamefont
  {Peccei}}\ and\ \bibinfo {author} {\bibfnamefont {H.~R.}\ \bibnamefont
  {Quinn}},\ }\href {\doibase 10.1103/PhysRevD.16.1791} {\bibfield  {journal}
  {\bibinfo  {journal} {Phys. Rev. D}\ }\textbf {\bibinfo {volume} {16}},\
  \bibinfo {pages} {1791} (\bibinfo {year} {1977}{\natexlab{b}})}\BibitemShut
  {NoStop}%
\bibitem [{\citenamefont {Weinberg}(1978)}]{Weinberg:1977ma}%
  \BibitemOpen
  \bibfield  {author} {\bibinfo {author} {\bibfnamefont {S.}~\bibnamefont
  {Weinberg}},\ }\href {\doibase 10.1103/PhysRevLett.40.223} {\bibfield
  {journal} {\bibinfo  {journal} {Phys. Rev. Lett.}\ }\textbf {\bibinfo
  {volume} {40}},\ \bibinfo {pages} {223} (\bibinfo {year} {1978})}\BibitemShut
  {NoStop}%
\bibitem [{\citenamefont {Wilczek}(1978)}]{Wilczek:1977pj}%
  \BibitemOpen
  \bibfield  {author} {\bibinfo {author} {\bibfnamefont {F.}~\bibnamefont
  {Wilczek}},\ }\href {\doibase 10.1103/PhysRevLett.40.279} {\bibfield
  {journal} {\bibinfo  {journal} {Phys. Rev. Lett.}\ }\textbf {\bibinfo
  {volume} {40}},\ \bibinfo {pages} {279} (\bibinfo {year} {1978})}\BibitemShut
  {NoStop}%
\bibitem [{\citenamefont {Tanabashi}\ \emph {et~al.}(2018)\citenamefont
  {Tanabashi} \emph {et~al.}}]{Tanabashi:2018oca}%
  \BibitemOpen
  \bibfield  {author} {\bibinfo {author} {\bibfnamefont {M.}~\bibnamefont
  {Tanabashi}} \emph {et~al.} (\bibinfo {collaboration} {Particle Data
  Group}),\ }\href {\doibase 10.1103/PhysRevD.98.030001} {\bibfield  {journal}
  {\bibinfo  {journal} {Phys. Rev.}\ }\textbf {\bibinfo {volume} {D98}},\
  \bibinfo {pages} {030001} (\bibinfo {year} {2018})}\BibitemShut {NoStop}%
\bibitem [{\citenamefont {Kim}(1979)}]{Kim:1979if}%
  \BibitemOpen
  \bibfield  {author} {\bibinfo {author} {\bibfnamefont {J.~E.}\ \bibnamefont
  {Kim}},\ }\href {\doibase 10.1103/PhysRevLett.43.103} {\bibfield  {journal}
  {\bibinfo  {journal} {Phys. Rev. Lett.}\ }\textbf {\bibinfo {volume} {43}},\
  \bibinfo {pages} {103} (\bibinfo {year} {1979})}\BibitemShut {NoStop}%
\bibitem [{\citenamefont {Shifman}\ \emph {et~al.}(1980)\citenamefont
  {Shifman}, \citenamefont {Vainshtein},\ and\ \citenamefont
  {Zakharov}}]{Shifman:1979if}%
  \BibitemOpen
  \bibfield  {author} {\bibinfo {author} {\bibfnamefont {M.~A.}\ \bibnamefont
  {Shifman}}, \bibinfo {author} {\bibfnamefont {A.~I.}\ \bibnamefont
  {Vainshtein}}, \ and\ \bibinfo {author} {\bibfnamefont {V.~I.}\ \bibnamefont
  {Zakharov}},\ }\href {\doibase 10.1016/0550-3213(80)90209-6} {\bibfield
  {journal} {\bibinfo  {journal} {Nucl. Phys. B}\ }\textbf {\bibinfo {volume}
  {166}},\ \bibinfo {pages} {493} (\bibinfo {year} {1980})}\BibitemShut
  {NoStop}%
\bibitem [{\citenamefont {Dine}\ \emph {et~al.}(1981)\citenamefont {Dine},
  \citenamefont {Fischler},\ and\ \citenamefont {Srednicki}}]{Dine:1981rt}%
  \BibitemOpen
  \bibfield  {author} {\bibinfo {author} {\bibfnamefont {M.}~\bibnamefont
  {Dine}}, \bibinfo {author} {\bibfnamefont {W.}~\bibnamefont {Fischler}}, \
  and\ \bibinfo {author} {\bibfnamefont {M.}~\bibnamefont {Srednicki}},\ }\href
  {\doibase 10.1016/0370-2693(81)90590-6} {\bibfield  {journal} {\bibinfo
  {journal} {Phys. Lett. B}\ }\textbf {\bibinfo {volume} {104}},\ \bibinfo
  {pages} {199} (\bibinfo {year} {1981})}\BibitemShut {NoStop}%
\bibitem [{\citenamefont {Zhitnitsky}(1980)}]{Zhitnitsky:1980tq}%
  \BibitemOpen
  \bibfield  {author} {\bibinfo {author} {\bibfnamefont {A.~R.}\ \bibnamefont
  {Zhitnitsky}},\ }\href@noop {} {\bibfield  {journal} {\bibinfo  {journal}
  {Sov. J. Nucl. Phys.}\ }\textbf {\bibinfo {volume} {31}},\ \bibinfo {pages}
  {260} (\bibinfo {year} {1980})}\BibitemShut {NoStop}%
\bibitem [{\citenamefont {Sikivie}(1983)}]{Sikivie:1983ip}%
  \BibitemOpen
  \bibfield  {author} {\bibinfo {author} {\bibfnamefont {P.}~\bibnamefont
  {Sikivie}},\ }\href {\doibase 10.1103/PhysRevLett.51.1415} {\bibfield
  {journal} {\bibinfo  {journal} {Phys. Rev. Lett.}\ }\textbf {\bibinfo
  {volume} {51}},\ \bibinfo {pages} {1415} (\bibinfo {year} {1983})},\ \bibinfo
  {note} {[Erratum: Phys.Rev.Lett. 52, 695 (1984)]}\BibitemShut {NoStop}%
\bibitem [{\citenamefont {Anselm}(1985)}]{Anselm:1985obz}%
  \BibitemOpen
  \bibfield  {author} {\bibinfo {author} {\bibfnamefont {A.~A.}\ \bibnamefont
  {Anselm}},\ }\href@noop {} {\bibfield  {journal} {\bibinfo  {journal} {Yad.
  Fiz.}\ }\textbf {\bibinfo {volume} {42}},\ \bibinfo {pages} {1480} (\bibinfo
  {year} {1985})}\BibitemShut {NoStop}%
\bibitem [{\citenamefont {Van~Bibber}\ \emph {et~al.}(1987)\citenamefont
  {Van~Bibber}, \citenamefont {Dagdeviren}, \citenamefont {Koonin},
  \citenamefont {Kerman},\ and\ \citenamefont {Nelson}}]{VanBibber:1987rq}%
  \BibitemOpen
  \bibfield  {author} {\bibinfo {author} {\bibfnamefont {K.}~\bibnamefont
  {Van~Bibber}}, \bibinfo {author} {\bibfnamefont {N.~R.}\ \bibnamefont
  {Dagdeviren}}, \bibinfo {author} {\bibfnamefont {S.~E.}\ \bibnamefont
  {Koonin}}, \bibinfo {author} {\bibfnamefont {A.}~\bibnamefont {Kerman}}, \
  and\ \bibinfo {author} {\bibfnamefont {H.~N.}\ \bibnamefont {Nelson}},\
  }\href {\doibase 10.1103/PhysRevLett.59.759} {\bibfield  {journal} {\bibinfo
  {journal} {Phys. Rev. Lett.}\ }\textbf {\bibinfo {volume} {59}},\ \bibinfo
  {pages} {759} (\bibinfo {year} {1987})}\BibitemShut {NoStop}%
\bibitem [{\citenamefont {Ehret}\ \emph {et~al.}(2009)\citenamefont {Ehret}
  \emph {et~al.}}]{ALPS:2009des}%
  \BibitemOpen
  \bibfield  {author} {\bibinfo {author} {\bibfnamefont {K.}~\bibnamefont
  {Ehret}} \emph {et~al.} (\bibinfo {collaboration} {ALPS}),\ }\href {\doibase
  10.1016/j.nima.2009.10.102} {\bibfield  {journal} {\bibinfo  {journal} {Nucl.
  Instrum. Meth. A}\ }\textbf {\bibinfo {volume} {612}},\ \bibinfo {pages} {83}
  (\bibinfo {year} {2009})},\ \Eprint {http://arxiv.org/abs/0905.4159}
  {arXiv:0905.4159 [physics.ins-det]} \BibitemShut {NoStop}%
\bibitem [{\citenamefont {B\"ahre}\ \emph {et~al.}(2013)\citenamefont {B\"ahre}
  \emph {et~al.}}]{Bahre:2013ywa}%
  \BibitemOpen
  \bibfield  {author} {\bibinfo {author} {\bibfnamefont {R.}~\bibnamefont
  {B\"ahre}} \emph {et~al.},\ }\href {\doibase 10.1088/1748-0221/8/09/T09001}
  {\bibfield  {journal} {\bibinfo  {journal} {JINST}\ }\textbf {\bibinfo
  {volume} {8}},\ \bibinfo {pages} {T09001} (\bibinfo {year} {2013})},\ \Eprint
  {http://arxiv.org/abs/1302.5647} {arXiv:1302.5647 [physics.ins-det]}
  \BibitemShut {NoStop}%
\bibitem [{\citenamefont {Ruoso}\ \emph {et~al.}(1992)\citenamefont {Ruoso}
  \emph {et~al.}}]{Ruoso:1992nx}%
  \BibitemOpen
  \bibfield  {author} {\bibinfo {author} {\bibfnamefont {G.}~\bibnamefont
  {Ruoso}} \emph {et~al.},\ }\href {\doibase 10.1007/BF01474722} {\bibfield
  {journal} {\bibinfo  {journal} {Z. Phys. C}\ }\textbf {\bibinfo {volume}
  {56}},\ \bibinfo {pages} {505} (\bibinfo {year} {1992})}\BibitemShut
  {NoStop}%
\bibitem [{\citenamefont {Cameron}\ \emph {et~al.}(1993)\citenamefont {Cameron}
  \emph {et~al.}}]{Cameron:1993mr}%
  \BibitemOpen
  \bibfield  {author} {\bibinfo {author} {\bibfnamefont {R.}~\bibnamefont
  {Cameron}} \emph {et~al.},\ }\href {\doibase 10.1103/PhysRevD.47.3707}
  {\bibfield  {journal} {\bibinfo  {journal} {Phys. Rev. D}\ }\textbf {\bibinfo
  {volume} {47}},\ \bibinfo {pages} {3707} (\bibinfo {year}
  {1993})}\BibitemShut {NoStop}%
\bibitem [{\citenamefont {Robilliard}\ \emph {et~al.}(2007)\citenamefont
  {Robilliard}, \citenamefont {Battesti}, \citenamefont {Fouche}, \citenamefont
  {Mauchain}, \citenamefont {Sautivet}, \citenamefont {Amiranoff},\ and\
  \citenamefont {Rizzo}}]{Robilliard:2007bq}%
  \BibitemOpen
  \bibfield  {author} {\bibinfo {author} {\bibfnamefont {C.}~\bibnamefont
  {Robilliard}}, \bibinfo {author} {\bibfnamefont {R.}~\bibnamefont
  {Battesti}}, \bibinfo {author} {\bibfnamefont {M.}~\bibnamefont {Fouche}},
  \bibinfo {author} {\bibfnamefont {J.}~\bibnamefont {Mauchain}}, \bibinfo
  {author} {\bibfnamefont {A.-M.}\ \bibnamefont {Sautivet}}, \bibinfo {author}
  {\bibfnamefont {F.}~\bibnamefont {Amiranoff}}, \ and\ \bibinfo {author}
  {\bibfnamefont {C.}~\bibnamefont {Rizzo}},\ }\href {\doibase
  10.1103/PhysRevLett.99.190403} {\bibfield  {journal} {\bibinfo  {journal}
  {Phys. Rev. Lett.}\ }\textbf {\bibinfo {volume} {99}},\ \bibinfo {pages}
  {190403} (\bibinfo {year} {2007})},\ \Eprint {http://arxiv.org/abs/0707.1296}
  {arXiv:0707.1296 [hep-ex]} \BibitemShut {NoStop}%
\bibitem [{\citenamefont {Chou}\ \emph {et~al.}(2008)\citenamefont {Chou},
  \citenamefont {Wester}, \citenamefont {Baumbaugh}, \citenamefont {Gustafson},
  \citenamefont {Irizarry-Valle}, \citenamefont {Mazur}, \citenamefont
  {Steffen}, \citenamefont {Tomlin}, \citenamefont {Yang},\ and\ \citenamefont
  {Yoo}}]{GammeVT-969:2007pci}%
  \BibitemOpen
  \bibfield  {author} {\bibinfo {author} {\bibfnamefont {A.~S.}\ \bibnamefont
  {Chou}}, \bibinfo {author} {\bibfnamefont {W.~C.}\ \bibnamefont {Wester},
  \bibfnamefont {III}}, \bibinfo {author} {\bibfnamefont {A.}~\bibnamefont
  {Baumbaugh}}, \bibinfo {author} {\bibfnamefont {H.~R.}\ \bibnamefont
  {Gustafson}}, \bibinfo {author} {\bibfnamefont {Y.}~\bibnamefont
  {Irizarry-Valle}}, \bibinfo {author} {\bibfnamefont {P.~O.}\ \bibnamefont
  {Mazur}}, \bibinfo {author} {\bibfnamefont {J.~H.}\ \bibnamefont {Steffen}},
  \bibinfo {author} {\bibfnamefont {R.}~\bibnamefont {Tomlin}}, \bibinfo
  {author} {\bibfnamefont {X.}~\bibnamefont {Yang}}, \ and\ \bibinfo {author}
  {\bibfnamefont {J.}~\bibnamefont {Yoo}} (\bibinfo {collaboration} {GammeV
  (T-969)}),\ }\href {\doibase 10.1103/PhysRevLett.100.080402} {\bibfield
  {journal} {\bibinfo  {journal} {Phys. Rev. Lett.}\ }\textbf {\bibinfo
  {volume} {100}},\ \bibinfo {pages} {080402} (\bibinfo {year} {2008})},\
  \Eprint {http://arxiv.org/abs/0710.3783} {arXiv:0710.3783 [hep-ex]}
  \BibitemShut {NoStop}%
\bibitem [{\citenamefont {Afanasev}\ \emph {et~al.}(2008)\citenamefont
  {Afanasev}, \citenamefont {Baker}, \citenamefont {Beard}, \citenamefont
  {Biallas}, \citenamefont {Boyce}, \citenamefont {Minarni}, \citenamefont
  {Ramdon}, \citenamefont {Shinn},\ and\ \citenamefont
  {Slocum}}]{Afanasev:2008jt}%
  \BibitemOpen
  \bibfield  {author} {\bibinfo {author} {\bibfnamefont {A.}~\bibnamefont
  {Afanasev}}, \bibinfo {author} {\bibfnamefont {O.~K.}\ \bibnamefont {Baker}},
  \bibinfo {author} {\bibfnamefont {K.~B.}\ \bibnamefont {Beard}}, \bibinfo
  {author} {\bibfnamefont {G.}~\bibnamefont {Biallas}}, \bibinfo {author}
  {\bibfnamefont {J.}~\bibnamefont {Boyce}}, \bibinfo {author} {\bibfnamefont
  {M.}~\bibnamefont {Minarni}}, \bibinfo {author} {\bibfnamefont
  {R.}~\bibnamefont {Ramdon}}, \bibinfo {author} {\bibfnamefont
  {M.}~\bibnamefont {Shinn}}, \ and\ \bibinfo {author} {\bibfnamefont
  {P.}~\bibnamefont {Slocum}},\ }\href {\doibase
  10.1103/PhysRevLett.101.120401} {\bibfield  {journal} {\bibinfo  {journal}
  {Phys. Rev. Lett.}\ }\textbf {\bibinfo {volume} {101}},\ \bibinfo {pages}
  {120401} (\bibinfo {year} {2008})},\ \Eprint {http://arxiv.org/abs/0806.2631}
  {arXiv:0806.2631 [hep-ex]} \BibitemShut {NoStop}%
\bibitem [{\citenamefont {Ballou}\ \emph {et~al.}(2015)\citenamefont {Ballou}
  \emph {et~al.}}]{OSQAR:2015qdv}%
  \BibitemOpen
  \bibfield  {author} {\bibinfo {author} {\bibfnamefont {R.}~\bibnamefont
  {Ballou}} \emph {et~al.} (\bibinfo {collaboration} {OSQAR}),\ }\href
  {\doibase 10.1103/PhysRevD.92.092002} {\bibfield  {journal} {\bibinfo
  {journal} {Phys. Rev. D}\ }\textbf {\bibinfo {volume} {92}},\ \bibinfo
  {pages} {092002} (\bibinfo {year} {2015})},\ \Eprint
  {http://arxiv.org/abs/1506.08082} {arXiv:1506.08082 [hep-ex]} \BibitemShut
  {NoStop}%
\bibitem [{\citenamefont {Inada}\ \emph {et~al.}(2017)\citenamefont {Inada}
  \emph {et~al.}}]{Inada:2016jzh}%
  \BibitemOpen
  \bibfield  {author} {\bibinfo {author} {\bibfnamefont {T.}~\bibnamefont
  {Inada}} \emph {et~al.},\ }\href {\doibase 10.1103/PhysRevLett.118.071803}
  {\bibfield  {journal} {\bibinfo  {journal} {Phys. Rev. Lett.}\ }\textbf
  {\bibinfo {volume} {118}},\ \bibinfo {pages} {071803} (\bibinfo {year}
  {2017})},\ \Eprint {http://arxiv.org/abs/1609.05425} {arXiv:1609.05425
  [hep-ex]} \BibitemShut {NoStop}%
\bibitem [{\citenamefont {Gai}\ \emph {et~al.}(2018)\citenamefont {Gai} \emph
  {et~al.}}]{positronSource}%
  \BibitemOpen
  \bibfield  {author} {\bibinfo {author} {\bibfnamefont {W.}~\bibnamefont
  {Gai}} \emph {et~al.} (\bibinfo {collaboration} {Positron Working Group}),\
  }\href {https://edmsdirect.desy.de/item/D00000001165115} {\enquote {\bibinfo
  {title} {{Report on the ILC Positron Source}},}\ } (\bibinfo {year}
  {2018})\BibitemShut {NoStop}%
\bibitem [{\citenamefont {Inoue}\ \emph {et~al.}(2002)\citenamefont {Inoue},
  \citenamefont {Namba}, \citenamefont {Moriyama}, \citenamefont {Minowa},
  \citenamefont {Takasu}, \citenamefont {Horiuchi},\ and\ \citenamefont
  {Yamamoto}}]{Inoue:2002qy}%
  \BibitemOpen
  \bibfield  {author} {\bibinfo {author} {\bibfnamefont {Y.}~\bibnamefont
  {Inoue}}, \bibinfo {author} {\bibfnamefont {T.}~\bibnamefont {Namba}},
  \bibinfo {author} {\bibfnamefont {S.}~\bibnamefont {Moriyama}}, \bibinfo
  {author} {\bibfnamefont {M.}~\bibnamefont {Minowa}}, \bibinfo {author}
  {\bibfnamefont {Y.}~\bibnamefont {Takasu}}, \bibinfo {author} {\bibfnamefont
  {T.}~\bibnamefont {Horiuchi}}, \ and\ \bibinfo {author} {\bibfnamefont
  {A.}~\bibnamefont {Yamamoto}},\ }\href {\doibase
  10.1016/S0370-2693(02)01822-1} {\bibfield  {journal} {\bibinfo  {journal}
  {Phys. Lett. B}\ }\textbf {\bibinfo {volume} {536}},\ \bibinfo {pages} {18}
  (\bibinfo {year} {2002})},\ \Eprint {http://arxiv.org/abs/astro-ph/0204388}
  {arXiv:astro-ph/0204388} \BibitemShut {NoStop}%
\bibitem [{\citenamefont {Arik}\ \emph {et~al.}(2009)\citenamefont {Arik} \emph
  {et~al.}}]{CAST_He4}%
  \BibitemOpen
  \bibfield  {author} {\bibinfo {author} {\bibfnamefont {E.}~\bibnamefont
  {Arik}} \emph {et~al.} (\bibinfo {collaboration} {CAST}),\ }\href {\doibase
  10.1088/1475-7516/2009/02/008} {\bibfield  {journal} {\bibinfo  {journal}
  {JCAP}\ }\textbf {\bibinfo {volume} {02}},\ \bibinfo {pages} {008} (\bibinfo
  {year} {2009})},\ \Eprint {http://arxiv.org/abs/0810.4482} {arXiv:0810.4482
  [hep-ex]} \BibitemShut {NoStop}%
\bibitem [{\citenamefont {Arik}\ \emph {et~al.}(2014)\citenamefont {Arik} \emph
  {et~al.}}]{CAST_He3}%
  \BibitemOpen
  \bibfield  {author} {\bibinfo {author} {\bibfnamefont {M.}~\bibnamefont
  {Arik}} \emph {et~al.} (\bibinfo {collaboration} {CAST}),\ }\href {\doibase
  10.1103/PhysRevLett.112.091302} {\bibfield  {journal} {\bibinfo  {journal}
  {Phys. Rev. Lett.}\ }\textbf {\bibinfo {volume} {112}},\ \bibinfo {pages}
  {091302} (\bibinfo {year} {2014})},\ \Eprint {http://arxiv.org/abs/1307.1985}
  {arXiv:1307.1985 [hep-ex]} \BibitemShut {NoStop}%
\bibitem [{\citenamefont {Adolphsen}\ \emph {et~al.}(2013)\citenamefont
  {Adolphsen} \emph {et~al.}}]{Adolphsen:2013kya}%
  \BibitemOpen
  \bibfield  {author} {\bibinfo {author} {\bibfnamefont {C.}~\bibnamefont
  {Adolphsen}} \emph {et~al.},\ }\href@noop {} {\  (\bibinfo {year} {2013})},\
  \Eprint {http://arxiv.org/abs/1306.6328} {arXiv:1306.6328 [physics.acc-ph]}
  \BibitemShut {NoStop}%
\bibitem [{\citenamefont {Povinec}\ \emph {et~al.}(2008)\citenamefont
  {Povinec}, \citenamefont {Betti}, \citenamefont {Jull},\ and\ \citenamefont
  {Vojtyla}}]{b8e9aa681f8b4f3291a77ffb2650033d}%
  \BibitemOpen
  \bibfield  {author} {\bibinfo {author} {\bibfnamefont {P.}~\bibnamefont
  {Povinec}}, \bibinfo {author} {\bibfnamefont {M.}~\bibnamefont {Betti}},
  \bibinfo {author} {\bibfnamefont {A.}~\bibnamefont {Jull}}, \ and\ \bibinfo
  {author} {\bibfnamefont {P.}~\bibnamefont {Vojtyla}},\ }\href
  {http://www.physics.sk/aps/pub.php?y=2008&pub=aps-08-01} {\bibfield
  {journal} {\bibinfo  {journal} {Acta Physica Slovaca}\ }\textbf {\bibinfo
  {volume} {58}},\ \bibinfo {pages} {1} (\bibinfo {year} {2008})}\BibitemShut
  {NoStop}%
\bibitem [{\citenamefont {Crespi}\ \emph {et~al.}(2010)\citenamefont {Crespi},
  \citenamefont {Vandone}, \citenamefont {Brambilla}, \citenamefont {Camera},
  \citenamefont {Million}, \citenamefont {Riboldi},\ and\ \citenamefont
  {Wieland}}]{509}%
  \BibitemOpen
  \bibfield  {author} {\bibinfo {author} {\bibfnamefont {F.}~\bibnamefont
  {Crespi}}, \bibinfo {author} {\bibfnamefont {V.}~\bibnamefont {Vandone}},
  \bibinfo {author} {\bibfnamefont {S.}~\bibnamefont {Brambilla}}, \bibinfo
  {author} {\bibfnamefont {F.}~\bibnamefont {Camera}}, \bibinfo {author}
  {\bibfnamefont {B.}~\bibnamefont {Million}}, \bibinfo {author} {\bibfnamefont
  {S.}~\bibnamefont {Riboldi}}, \ and\ \bibinfo {author} {\bibfnamefont
  {O.}~\bibnamefont {Wieland}},\ }\href {\doibase 10.1016/j.nima.2010.02.273}
  {\bibfield  {journal} {\bibinfo  {journal} {Nuclear Instruments and Methods
  in Physics Research Section A: Accelerators, Spectrometers, Detectors and
  Associated Equipment}\ }\textbf {\bibinfo {volume} {620}},\ \bibinfo {pages}
  {299} (\bibinfo {year} {2010})},\ \bibinfo {note} {received 23 December 2009,
  Revised 24 February 2010, Accepted 27 February 2010, Available online 15
  March 2010.}\BibitemShut {Stop}%
\bibitem [{\citenamefont {Caputo}\ \emph {et~al.}(2019)\citenamefont {Caputo}
  \emph {et~al.}}]{AMEGO:2019gny}%
  \BibitemOpen
  \bibfield  {author} {\bibinfo {author} {\bibfnamefont {R.}~\bibnamefont
  {Caputo}} \emph {et~al.} (\bibinfo {collaboration} {AMEGO}),\ }\href@noop {}
  {\  (\bibinfo {year} {2019})},\ \Eprint {http://arxiv.org/abs/1907.07558}
  {arXiv:1907.07558 [astro-ph.IM]} \BibitemShut {NoStop}%
\bibitem [{\citenamefont {De~Angelis}\ \emph {et~al.}(2017)\citenamefont
  {De~Angelis} \emph {et~al.}}]{e-ASTROGAM:2016bph}%
  \BibitemOpen
  \bibfield  {author} {\bibinfo {author} {\bibfnamefont {A.}~\bibnamefont
  {De~Angelis}} \emph {et~al.} (\bibinfo {collaboration} {e-ASTROGAM}),\ }\href
  {\doibase 10.1007/s10686-017-9533-6} {\bibfield  {journal} {\bibinfo
  {journal} {Exper. Astron.}\ }\textbf {\bibinfo {volume} {44}},\ \bibinfo
  {pages} {25} (\bibinfo {year} {2017})},\ \Eprint
  {http://arxiv.org/abs/1611.02232} {arXiv:1611.02232 [astro-ph.HE]}
  \BibitemShut {NoStop}%
\bibitem [{\citenamefont {Tavani}\ \emph {et~al.}(2018)\citenamefont {Tavani}
  \emph {et~al.}}]{e-ASTROGAM:2017pxr}%
  \BibitemOpen
  \bibfield  {author} {\bibinfo {author} {\bibfnamefont {M.}~\bibnamefont
  {Tavani}} \emph {et~al.} (\bibinfo {collaboration} {e-ASTROGAM}),\ }\href
  {\doibase 10.1016/j.jheap.2018.07.001} {\bibfield  {journal} {\bibinfo
  {journal} {JHEAp}\ }\textbf {\bibinfo {volume} {19}},\ \bibinfo {pages} {1}
  (\bibinfo {year} {2018})},\ \Eprint {http://arxiv.org/abs/1711.01265}
  {arXiv:1711.01265 [astro-ph.HE]} \BibitemShut {NoStop}%
\bibitem [{\citenamefont {Alharbi}\ \emph
  {et~al.}(2021{\natexlab{a}})\citenamefont {Alharbi}, \citenamefont {Riemann},
  \citenamefont {Alrashdi}, \citenamefont {Moortgat-Pick}, \citenamefont
  {Ushakov},\ and\ \citenamefont {Sievers}}]{Alharbi:2021okl}%
  \BibitemOpen
  \bibfield  {author} {\bibinfo {author} {\bibfnamefont {K.}~\bibnamefont
  {Alharbi}}, \bibinfo {author} {\bibfnamefont {S.}~\bibnamefont {Riemann}},
  \bibinfo {author} {\bibfnamefont {A.}~\bibnamefont {Alrashdi}}, \bibinfo
  {author} {\bibfnamefont {G.}~\bibnamefont {Moortgat-Pick}}, \bibinfo {author}
  {\bibfnamefont {A.}~\bibnamefont {Ushakov}}, \ and\ \bibinfo {author}
  {\bibfnamefont {P.}~\bibnamefont {Sievers}},\ }in\ \href@noop {} {\emph
  {\bibinfo {booktitle} {{International Workshop on Future Linear
  Colliders}}}}\ (\bibinfo {year} {2021})\ \Eprint
  {http://arxiv.org/abs/2106.00074} {arXiv:2106.00074 [physics.acc-ph]}
  \BibitemShut {NoStop}%
\bibitem [{\citenamefont {Alharbi}\ \emph
  {et~al.}(2021{\natexlab{b}})\citenamefont {Alharbi}, \citenamefont
  {Alrashdi}, \citenamefont {Moortgat-Pick}, \citenamefont {Riemann},
  \citenamefont {Sievers},\ and\ \citenamefont {Ushakov}}]{Alharbi:2021ctf}%
  \BibitemOpen
  \bibfield  {author} {\bibinfo {author} {\bibfnamefont {K.}~\bibnamefont
  {Alharbi}}, \bibinfo {author} {\bibfnamefont {A.}~\bibnamefont {Alrashdi}},
  \bibinfo {author} {\bibfnamefont {G.}~\bibnamefont {Moortgat-Pick}}, \bibinfo
  {author} {\bibfnamefont {S.}~\bibnamefont {Riemann}}, \bibinfo {author}
  {\bibfnamefont {P.}~\bibnamefont {Sievers}}, \ and\ \bibinfo {author}
  {\bibfnamefont {A.}~\bibnamefont {Ushakov}},\ }\href {\doibase
  10.18429/JACoW-IPAC2021-THPAB041} {\bibfield  {journal} {\bibinfo  {journal}
  {JACoW}\ }\textbf {\bibinfo {volume} {IPAC2021}},\ \bibinfo {pages}
  {THPAB041} (\bibinfo {year} {2021}{\natexlab{b}})}\BibitemShut {NoStop}%
\bibitem [{\citenamefont {Di~Luzio}\ \emph
  {et~al.}(2017{\natexlab{a}})\citenamefont {Di~Luzio}, \citenamefont
  {Mescia},\ and\ \citenamefont {Nardi}}]{DiLuzio:2016sbl}%
  \BibitemOpen
  \bibfield  {author} {\bibinfo {author} {\bibfnamefont {L.}~\bibnamefont
  {Di~Luzio}}, \bibinfo {author} {\bibfnamefont {F.}~\bibnamefont {Mescia}}, \
  and\ \bibinfo {author} {\bibfnamefont {E.}~\bibnamefont {Nardi}},\ }\href
  {\doibase 10.1103/PhysRevLett.118.031801} {\bibfield  {journal} {\bibinfo
  {journal} {Phys. Rev. Lett.}\ }\textbf {\bibinfo {volume} {118}},\ \bibinfo
  {pages} {031801} (\bibinfo {year} {2017}{\natexlab{a}})},\ \Eprint
  {http://arxiv.org/abs/1610.07593} {arXiv:1610.07593 [hep-ph]} \BibitemShut
  {NoStop}%
\bibitem [{\citenamefont {Di~Luzio}\ \emph
  {et~al.}(2017{\natexlab{b}})\citenamefont {Di~Luzio}, \citenamefont
  {Mescia},\ and\ \citenamefont {Nardi}}]{DiLuzio:2017pfr}%
  \BibitemOpen
  \bibfield  {author} {\bibinfo {author} {\bibfnamefont {L.}~\bibnamefont
  {Di~Luzio}}, \bibinfo {author} {\bibfnamefont {F.}~\bibnamefont {Mescia}}, \
  and\ \bibinfo {author} {\bibfnamefont {E.}~\bibnamefont {Nardi}},\ }\href
  {\doibase 10.1103/PhysRevD.96.075003} {\bibfield  {journal} {\bibinfo
  {journal} {Phys. Rev. D}\ }\textbf {\bibinfo {volume} {96}},\ \bibinfo
  {pages} {075003} (\bibinfo {year} {2017}{\natexlab{b}})},\ \Eprint
  {http://arxiv.org/abs/1705.05370} {arXiv:1705.05370 [hep-ph]} \BibitemShut
  {NoStop}%
\bibitem [{\citenamefont {Inada}\ \emph {et~al.}(2013)\citenamefont {Inada},
  \citenamefont {Namba}, \citenamefont {Asai}, \citenamefont {Kobayashi},
  \citenamefont {Tanaka}, \citenamefont {Tamasaku}, \citenamefont {Sawada},\
  and\ \citenamefont {Ishikawa}}]{Inada:2013tx}%
  \BibitemOpen
  \bibfield  {author} {\bibinfo {author} {\bibfnamefont {T.}~\bibnamefont
  {Inada}}, \bibinfo {author} {\bibfnamefont {T.}~\bibnamefont {Namba}},
  \bibinfo {author} {\bibfnamefont {S.}~\bibnamefont {Asai}}, \bibinfo {author}
  {\bibfnamefont {T.}~\bibnamefont {Kobayashi}}, \bibinfo {author}
  {\bibfnamefont {Y.}~\bibnamefont {Tanaka}}, \bibinfo {author} {\bibfnamefont
  {K.}~\bibnamefont {Tamasaku}}, \bibinfo {author} {\bibfnamefont
  {K.}~\bibnamefont {Sawada}}, \ and\ \bibinfo {author} {\bibfnamefont
  {T.}~\bibnamefont {Ishikawa}},\ }\href {\doibase
  10.1016/j.physletb.2013.04.033} {\bibfield  {journal} {\bibinfo  {journal}
  {Phys. Lett. B}\ }\textbf {\bibinfo {volume} {722}},\ \bibinfo {pages} {301}
  (\bibinfo {year} {2013})},\ \Eprint {http://arxiv.org/abs/1301.6557}
  {arXiv:1301.6557 [physics.ins-det]} \BibitemShut {NoStop}%
\bibitem [{\citenamefont {Akhmedov}\ and\ \citenamefont
  {Smirnov}(2011)}]{Akhmedov:2010ua}%
  \BibitemOpen
  \bibfield  {author} {\bibinfo {author} {\bibfnamefont {E.~K.}\ \bibnamefont
  {Akhmedov}}\ and\ \bibinfo {author} {\bibfnamefont {A.~Y.}\ \bibnamefont
  {Smirnov}},\ }\href {\doibase 10.1007/s10701-011-9545-4} {\bibfield
  {journal} {\bibinfo  {journal} {Found. Phys.}\ }\textbf {\bibinfo {volume}
  {41}},\ \bibinfo {pages} {1279} (\bibinfo {year} {2011})},\ \Eprint
  {http://arxiv.org/abs/1008.2077} {arXiv:1008.2077 [hep-ph]} \BibitemShut
  {NoStop}%
\end{thebibliography}%

\end{document}